\documentclass[pra,letterpaper,aps,10pt,superscriptaddress,twocolumn,floatfix,showpacs]{revtex4-1}
\usepackage{amsmath,graphicx,amssymb,braket,xcolor,subfigure,upgreek,placeins,natbib,url,hyperref}
\usepackage{color}
\usepackage{microtype}

\newcommand{\fref}[1]{figure \ref{#1}}
\renewcommand{\subsubsection}[1]{\vspace{\baselineskip}\noindent\emph{#1}. --}

\begin{document}

\title{Laser noise imposed limitations of ensemble quantum metrology}
\author{D. Plankensteiner}
\affiliation{Institut f\"ur Theoretische Physik, Universit\"at Innsbruck, Technikerstrasse 21a, A-6020 Innsbruck, Austria}
\author{J. Schachenmayer}
\affiliation{JILA, NIST, Department of Physics, University of Colorado, 440 UCB, Boulder, CO 80309, USA}
\author{H. Ritsch}
\affiliation{Institut f\"ur Theoretische Physik, Universit\"at Innsbruck, Technikerstrasse 21a, A-6020 Innsbruck, Austria}
\author{C. Genes}
\affiliation{Institut f\"ur Theoretische Physik, Universit\"at Innsbruck, Technikerstrasse 21a, A-6020 Innsbruck, Austria}
\date{\today}
\affiliation{Vienna Center for Quantum Science and Technology, TU Wien-Atominstitut, Stadionallee 2, 1020 Vienna, Austria}
\email[Email address:]{claudiu.genes@uibk.ac.at‎}

\begin{abstract}
Laser noise is a decisive limiting factor in high precision spectroscopy of narrow lines using atomic ensembles. In an idealized Doppler and differential light shift free magic wavelength lattice configuration, it remains as one distinct principal limitation beyond collective atomic decay. In this work we study the limitations originating from laser phase and amplitude noise in an idealized Ramsey pulse interrogation scheme with uncorrelated atoms. Phase noise leads to a saturation of the frequency sensitivity with increasing atom number while amplitude noise implies a scaling $1/\sqrt{\tau}$ with $\tau$ being the interrogation time. We employ a technique using decoherence free subspaces first introduced in New J. Phys. \textbf{14}, 043011 (2012) which can restore the scaling with the square root of the inverse particle number $1/\sqrt{N}$. Similar results and improvements are obtained numerically for a Rabi spectroscopy setup.
\end{abstract}

\pacs{42.50.Ar, 42.50.Lc, 42.72.-g}

\maketitle

\section{Introduction}
Ultracold atoms in optical lattices~\cite{bloch2008many,lewenstein2007ultracold} can nowadays be almost perfectly controlled on the quantum level of motion and internal degrees of freedom. Using special magic wavelength configurations for alkaline earth atoms, the lattice introduces only minimal perturbations of the transition frequency~\cite{katori2011optical}. Such systems are perfectly suited candidates for state-of-the-art precision measurements. Optical lattice atomic clocks as cutting edge time standards are the most famous example of such exceedingly high resolution experiments~\cite{ludlow2015optical}. Recent endeavors have found what is yet the most accurate determination of time with a fractional uncertainty of $2\times 10^{-18}$~ \citep{nicholson2015systematic}. To achieve such a small total measurement error known noise sources (e.g. atomic collisions) had to be eliminated or corrected for by proper calibration. One remaining class of imprecisions is tied to the fairly high density of the atomic ensemble, which leads to radiative long range atom-atom interactions like collective decay (super-radiance~\cite{dicke1954coherence}) and dipole-dipole shifts~\cite{ficek2002entangled}. As they cannot be totally avoided, recent calculations targeted the suppression of atomic interactions~\cite{chang2004controlling,ostermann2013protected,ostermann2014protected} via optimized lattice geometries and symmetries.

A different but important uncertainty contribution for the clock is the finite linewidth of the reference laser used to interrogate the atomic ensemble. Despite tremendous efforts to construct ultra-stable reference cavities resulting in sub-Hertz oscillators, the randomly fluctuating phase of the laser still contributes significantly to the uncertainty budget in experiments~\cite{nicholson2015systematic}. This phase noise acts identically on all atoms and hence adds to the collective atomic dipole. Therefore the effect cannot be canceled by increasing the size of the ensemble. Mathematically, the effect of a noisy laser addressing the atomic ensemble leads to an exponential dephasing of the collective atomic dipole reducing the amount of information about the phase state of the atoms. Over the course of multiple measurements the phase noise aliases (Dick effect~\cite{dick1987local}) and degrades the stability of the clock.

First studies of the effects of laser noise on spectroscopy have been put forward already decades ago almost in parallel with the successful development of high resolution laser spectroscopy~\cite{wieman1976doppler}. Fluctuations in the laser phase and intensity lead to atomic population fluctuations limiting the spectroscopic resolution~\cite{hamilton1987saturation,haslwanter1988laser}. With the fast improvements in laser technology the linewidth of lasers soon became so small, that this effect could be largely ignored in typical setups. Only in the ultimate limit of clock transitions the linewidth is significantly smaller than available lasers. In this case the remaining fast fluctuations can be traced to thermal fluctuations in the mirror coatings while slow changes in the cavity length lead to a drift of the average oscillation frequency~\cite{ludlow2015optical}.

Numerous studies have analytically and numerically treated the effect of both phase and amplitude noise. These studies have been mostly focused on deriving spectroscopic precision limits for the case of multipartite entangled input states~\cite{jeske2014quantum,macieszczak2014bayesian,chaves2013noisy,brask2015improved,knysh2014true,szankowski2014parameter,dorner2012quantum}. Most of these approaches aimed to compute bounds for noisy metrology beyond the standard quantum limit. While it has generally been assumed that laser noise is especially detrimental to the typically fragile multipartite entangled states, significant limitations may occur already for 'classical' product input states.

In this paper we restrict our calculations to such product states in order to address current limitations of standard setups used in high-precision frequency detection. We derive scaling laws for the frequency sensitivity of both Ramsey and Rabi spectroscopy in the presence of collective laser noise. We map the stochastic laser-induced dynamics into an atomic master equation and show that amplitude noise has little impact on the precision. However, collective laser phase noise can lead to a complete saturation of the frequency sensitivity with the atom number (as shown for entangled states in Ref. \cite{dorner2012quantum}). We apply the proposal from Ref.~\cite{dorner2012quantum}, which uses decoherence free subspaces for the efficient suppression of phase noise. This is also reminiscent of the mechanism of quantum noise cancelation in atomic or optomechanical systems~\cite{ham2009establishing,caves2010coherent,caves2012evading,heurs2014coherent,hammerer2015trajectories,polzik2010quantum,clerk2013two,zhang2013back,vitali2016force}. Our scheme is analogous to the negative mass oscillator employed in the aforementioned optomechenical investigations. The suppression can be achieved by splitting the atomic ensemble into two sub-ensembles: the incident laser beam is also split and manipulated such that one sub-ensemble experiences the exact negative detuning (between the laser and the atomic transition frequency) of the other.

\section{Master equation for atomic dynamics in a noisy laser}

We consider an ensemble of identical two-level atoms with a transition frequency $\omega_0$. In a frame rotating with the laser frequency $\omega_l$ the ensemble is following free dynamics under the Hamiltonian ($\hbar=1$)
\begin{align}
H_0 &= \Omega S_z,
\end{align}
where $S_z=\sum_j \sigma_j^z/2$, $\sigma_j^z$ is the $z$ Pauli operator of the $j$-th atom and $\Omega=\omega_0-\omega_l$ is the detuning between the atomic transition and the laser frequency. The Hamiltonian describing optical excitation of this ensemble reads
\begin{align} \label{H_laser}
H_l &= 2\eta S_x,
\end{align}
where $\eta$ is the coherent pump strength and $S_x=\sum_j\sigma_j^x/2$. We assume that all atoms instantaneously feel any change of the laser phase, which is justified given that typical setups are of much smaller length than the coherence length of a standard laser~\cite{ye2012comparison}. Furthermore, assuming white noise (i.e. no correlations in time) we may describe the phase noise of a laser as collective dephasing of the atoms via the Lindblad term
\begin{align} \label{Ld_def}
\mathcal{L}_d[\rho] &= \frac{\gamma_d}{2}\left(2S_z\rho S_z - S_z^2\rho - \rho S_z^2\right).
\end{align}
Here, $\gamma_d$ is the strength of the dephasing at which the off-diagonal density matrix elements are damped out. The time dynamics for the atomic density matrix $\rho$ including laser phase noise are then described by the master equation
\begin{align} \label{master_eq1}
\dot{\rho} &= i\left[\rho,H\right] + \mathcal{L}_d[\rho],
\end{align}
where $H = H_0+H_l$. For details of the derivation of this master equation see Ref. \cite{dorner2012quantum} or Appendix A.

Analogously, if the laser has a noisy amplitude given by another white noise process and any change of amplitude instantly affects all atoms, we may describe the dynamics of the atomic ensemble by replacing the Lindblad term in the above master equation with (see also Appendix A)
\begin{align} \label{La_def}
\mathcal{L}_a[\rho] &= 2\gamma_a\left(2S_x\rho S_x - S_x^2\rho - \rho S_x^2\right),
\end{align}
where $\gamma_a$ governs the magnitude of the noise. Hence, amplitude noise can be interpreted as collective energy redistribution within the atomic ensemble.

Note, that one key point of this model is that all noise processes are of collective nature. This allows for simplified analytical and numerical treatment since the evolution described by the master equation takes place on the surface of the collective Bloch sphere, in a Hilbert space of dimension $N+1$.

\section{Effects of laser noise in Ramsey spectroscopy}

Ramsey spectroscopy~\cite{ramsey1950molecular} consists of two consecutive $\pi/2$-pulses applied to an atomic ensemble initially in the ground state. In between the two pulses the atoms are subjected to a period of free time evolution for a time $\tau$. After the second pulse the total population inversion of the atomic ensemble, which is proportional to the expectation value $\braket{S_z}$, is measured. This process is repeated for different laser frequencies (detunings) to retrieve the excitation as a function of the delay time and the detuning, $\braket{S_z}(\Omega,\tau)$. The $\pi/2$-pulses are assumed to be much faster than characteristic dephasing times and are therefore well approximated by rotations of the collective Bloch vector of the atoms about the y-axis of the Bloch sphere by an angle of $\pi/2$ (at exact resonance).

A figure of merit for the total frequency measurement precision is the so-called signal sensitivity $\delta\Omega$ \cite{wineland1994squeezed},
\begin{align} \label{sens}
\delta\Omega &= \min_{\Omega}\left[\frac{\Delta S_z(\Omega,\tau)}{|\partial_{\Omega}\braket{S_z}(\Omega,\tau)|}\right].
\end{align}
This quantity characterizes the minimal distinguishable frequency shift in the setup and thus gives the fundamental limit of accuracy with which one can match the laser frequency to the atomic transition frequency. The optimal operation points are those where the signal is most sensitive to changes in the detuning $\Omega$ while the signal quantum standard deviation $\Delta S_z=\sqrt{\braket{S_z^2}-\braket{S_z}^2}$ is minimized.

\subsection{Phase noise}

Let us first estimate the effect of the collective phase noise on the signal sensitivity obtained by interrogating an ensemble of $N$ atoms with a laser exhibiting phase noise. As previously mentioned, due to the collective nature of the Hamiltonian and the Liouvillian, all system dynamics take place in the symmetric subspace, i.e. the Dicke basis. The density matrix in this basis is
\begin{align} \label{rho_dicke}
\rho(t) &= \sum_{M,M'}\rho_{M,M'}(t)\ket{S,M}\bra{S,M'},
\end{align}
where $S=N/2$ and $M,M'=-S,-S+1,...,S$. Substituting Eq. \eqref{rho_dicke} into the master equation, i.e. Eq. \eqref{master_eq1}, we find differential equations for all density matrix elements of the form
\begin{align}
\dot{\rho}_{M,M'} &= \left[i\Omega (M'-M) -\frac{\gamma_d}{2}(M'-M)^2\right]\rho_{M,M'}.
\end{align}
Since the derivative of each density matrix element is proportional to the matrix element itself, an integration of the equation above is straightforward.
\begin{figure}[t]
\includegraphics[width=.8\columnwidth]{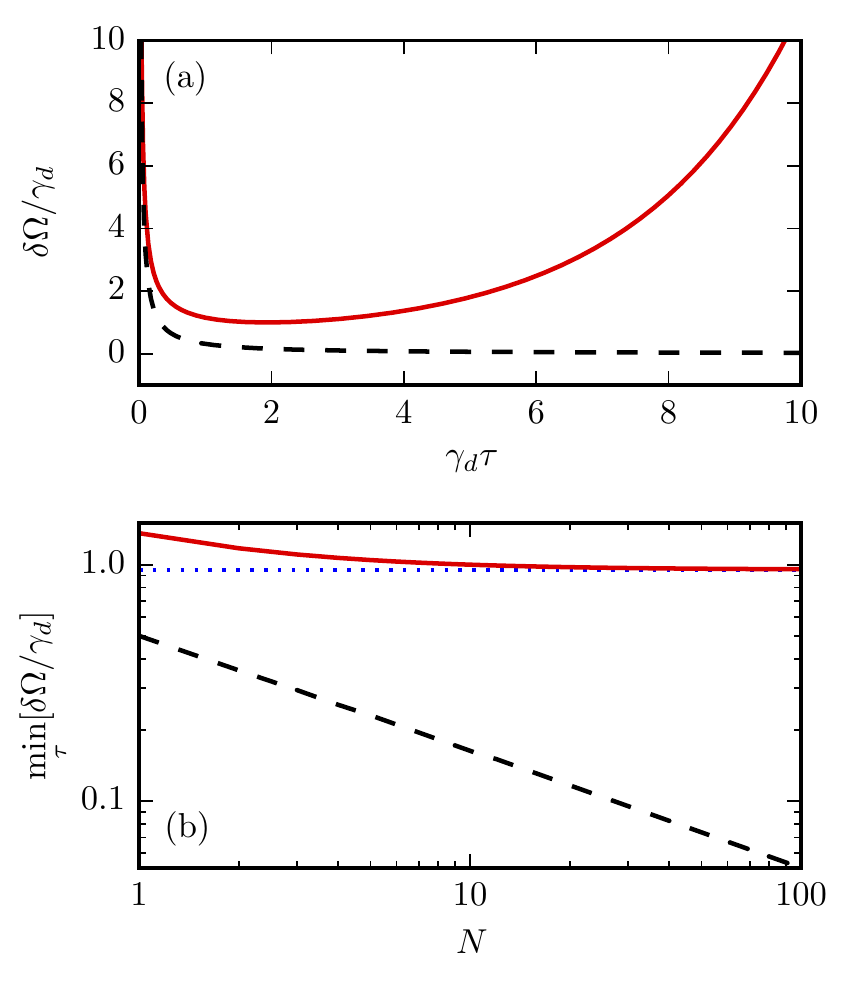}
\caption{\emph{Effects of phase noise}.  (a) The phase noise limited (red line) signal sensitivity is compared to the quantum projection limit (black dashed line) of a Ramsey measurement on $N=10$ atoms as a function of the interrogation time. The optimal time is $\tau_{opt}\approx 2/\gamma_d$. (b) The optimal signal sensitivity in the presence of phase noise saturates quickly with increasing $N$ approaching the lower bound (blue dots). This is compared to the projection noise limited curve scaling with $1/\sqrt{N}$ (for a fixed interrogation time $\tau_{opt}$) on a double-logarithmic scale.}
\label{fig1}
\end{figure}
The initial state for the free time evolution (the state after the first $\pi/2$-pulse) is described by the density matrix with the elements
\begin{align}
\rho_{M,M'}(0) &= \frac{1}{2^N}\left[\binom{N}{M}\binom{N}{M'}\right]^{\frac{1}{2}}.
\end{align}
Using the resulting solutions, we find that the expectation value $\braket{S_z}$ after the second $\pi/2$-pulse is
\begin{align} \label{signal}
\braket{S_z} &= \frac{N}{2}e^{-\gamma_d\tau/2}\cos{(\Omega\tau)}.
\end{align}
A somewhat more involved but nevertheless straightforward computation delivers the expression of the variance, which is proportional to the expectation value of the squared signal operator (for details see Appendix B),
\begin{align}
\label{variance}
\braket{S_z^2} &= \frac{N}{4}e^{-\gamma_d\tau}\left[N\sinh(\gamma_d\tau)+\cosh(\gamma_d\tau)\right].
\end{align}

Now the minimization with respect to $\Omega$ in Eq.~\eqref{sens} can be carried out. The minimum is found where the derivative of the signal $\partial_{\Omega}\braket{S_z}$ is extremal, i.e. $|\sin(\Omega\tau)|=1$. The final expression of the signal sensitivity for a standard Ramsey experiment is
\begin{align} \label{sens_sym}
\delta\Omega &= \frac{\sqrt{N\sinh(\gamma_d\tau)+\cosh(\gamma_d\tau)}}{\tau\sqrt{N}}.
\end{align}

An illustration of the temporal behavior of the sensitivity for $10$ atoms is given in~\fref{fig1}(a). In the absence of phase noise ($\gamma_d=0$) we recover the quantum projection noise limit. The most prominent characteristic is that already for a small number of atoms the sensitivity saturates, i.e. it omits its scaling with the atom number. The value which it saturates to is found by taking the limit $N\gg1$:
\begin{align}
\delta\Omega \approx \frac{\sqrt{\sinh(\gamma_d\tau)}}{\tau}.
\end{align}
The optimization over $\tau$ leads to a transcendental equation $\tanh(\gamma_d\tau/2)=\gamma_d\tau/2$ which can be numerically solved to indicate that the optimal sensitivity is simply limited by the bandwidth of the noise,
\begin{align} \label{sens_sym_min}
\min_\tau[\delta\Omega]\approx 0.951\gamma_d.
\end{align}
This behavior is depicted in \fref{fig1}(b), where the signal sensitivity optimized with respect to the interrogation time asymptotically approaches this lower bound. A similar bound that shows a saturation with the atom number $N$ was derived for GHZ-states in Ref.~\cite{dorner2012quantum}.

\subsection{Amplitude noise}

\begin{figure}[t]
\includegraphics[width=.8\columnwidth]{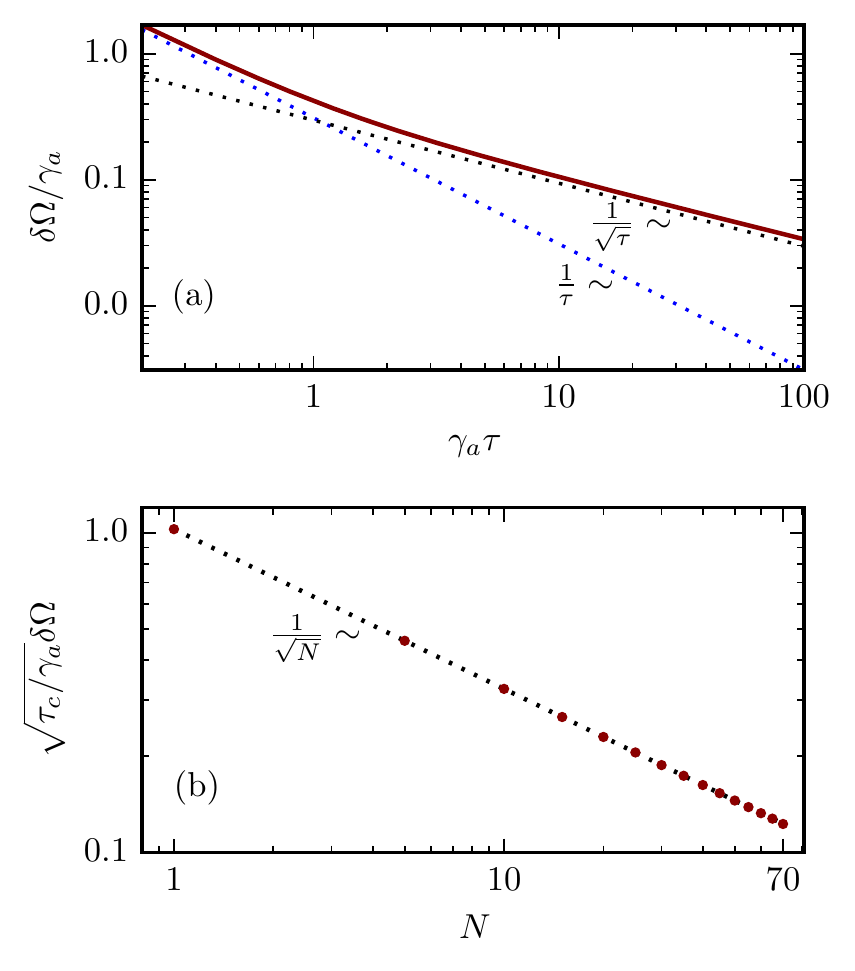}
\caption{\emph{Effects of amplitude noise}. (a) In order to illustrate the scaling with the interrogation time we have numerically computed the signal sensitivity in a standard Ramsey measurement performed on $N=10$ atoms. As can be seen in the double-logarithmic plot, the sensitivity scales like the quantum projection noise $1/\tau$ at short times and it transitions to a scaling with $1/\sqrt{\tau}$ for larger time scales. (b) The amplitude noise limit exhibits a scaling law of $1/\sqrt{N}$ with the atom number.}
\label{fig2}
\end{figure}

We now analyze the dynamics described by Eq.~\eqref{La_def} which includes the effects of amplitude noise. Using standard numerical methods, we can immediately uncover a particular limitation introduced by amplitude noise. The scaling of the sensitivity is modified from the typical $1/\tau$ to $1/\sqrt{\tau}$. This is illustrated in~\fref{fig2}(a) where the sensitivities obtained in both the degradation-free regime ($\gamma_a=0$) and for $\gamma_a> 0$ are compared. The crossover between the two scaling regimes is reached at around the inverse of the amplitude noise bandwidth $\tau\sim 1/\gamma_a$.

We now fix a duration of the free time evolution $\tau_c\gg\gamma_a^{-1}$ deep inside the $1/\sqrt{\tau}$ scaling regime and investigate the scaling of the sensitivity with increasing $N$. Numerical simulations [see~\fref{fig2}(b)] carried up to a fairly large number of atoms $N=70$ suggest that amplitude noise does not modify the scaling of the sensitivity with $N$,
\begin{align}
\delta\Omega\Big|_{\tau\gg\gamma_a^{-1}} &\propto \frac{1}{\sqrt{\tau N}}.
\end{align}

Under typical experimental conditions amplitude noise is orders of magnitude smaller than the phase noise, i.e. $\gamma_a\ll\gamma_d$. Therefore, as suggested by the analysis above, the first signature of laser noise is the saturation of the sensitivity with an increasing atom number occurring at times much smaller than the times where amplitude noise becomes important. \\
Let us note that this result agrees with the conclusion of Ref.~\cite{chaves2013noisy} where uncorrelated perpendicular noise (equivalent to independent amplitude noise) is shown to be far less detrimental than parallel noise (phase noise) for the spectroscopic resolution with entangled initial states.

\section{Circumventing the phase noise induced saturation via twin beam interrogation}

To overcome the saturation effect introduced by the noise associated with laser phase and frequency fluctuations, we employ a twin interrogation technique as in Ref.~\cite{dorner2012quantum}. This assumes the division of the $N$-atom ensemble into two separately addressable sub-ensembles of $N/2$ atoms each (with $N$ even and $S_z^{(1,2)}$ the corresponding population difference operators). Moreover, we set the detunings of the two ensembles opposite to each other leading to the Hamiltonian
\begin{align} \label{H_asym}
H_0^- &= \Omega S_z^{(1)} - \Omega S_z^{(2)},
\end{align}
as a replacement of the free evolution Hamiltonian in the standard Ramsey method. This constitutes an effective time-reversal operation as the sub-ensembles' associated dipoles rotate opposite to each other in time under the action of the same Hamiltonian. This cannot be simply realized by shifting the frequency of each ensemble upwards and downwards compared to the laser frequency but rather involves changing the sign of the laser frequency.

\begin{figure}[t]
\includegraphics[width=.8\columnwidth]{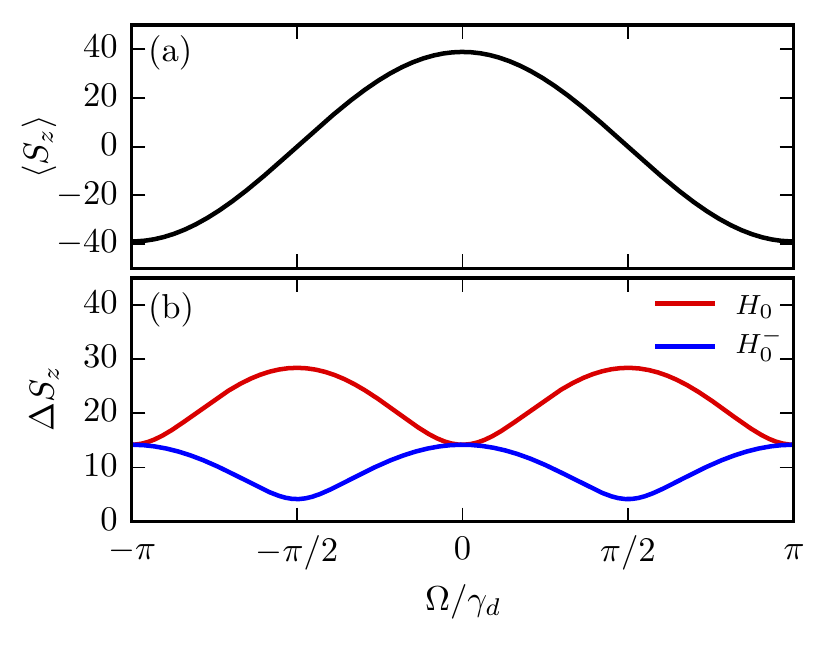}
\caption{\emph{The signal and its quantum standard deviation as a function of the detuning}. The central Ramsey fringe (a) and its standard deviation (b), for $N=100$ and interrogation time $\tau = 0.5/\gamma_d$. The deviation is significantly reduced at the steepest parts of the signal when using opposite detunings which allows for a more precise measurement of frequency.}
\label{fig5}
\end{figure}

The key point of this mechanism is that, while the systems evolve freely with $S_z^{(1)} - S_z^{(2)}$, the noise retains its collective nature $S_z^{(1)}+S_z^{(2)}$ according to Eq.~\eqref{Ld_def}. In a first step we notice that the difference in detunings in Eq.~\eqref{H_asym} breaks the symmetry such that the complete Hilbert space can no longer be described by a Dicke basis. Each sub-ensemble in itself though is restricted to its symmetric subspace, such that we can describe each ensemble with a Dicke basis. The complete basis is then just the product basis of the subspaces. We may write the total atomic density matrix as
\begin{align}
\rho(t) &= \sum_{\substack{m_1,m_2\\m_1',m_2'}}\rho_{m_1,m_2}^{m_1',m_2'}(t)\ket{s,m_1,m_2}\bra{s,m_1',m_2'},
\end{align}
where $s=N/4$ and each sum runs from $-s$ to $s$. The differential equations for the density matrix elements we obtain from the master equation now read
\begin{align}
\dot{\rho}_{m_1,m_2}^{m_1',m_2'} &= \bigg[i\Omega \left(m_1'-m_1\right) - i\Omega \left(m_2'-m_2\right)+
\notag \\
&- \frac{\gamma_d}{2}\left(m_1'+m_2'-(m_1+m_2)\right)^2\bigg]\rho_{m_1,m_2}^{m_1',m_2'}.
\end{align}
Their integration is straightforward and the density matrix at time $\tau$ can be derived (for details see Appendix B) for the initial conditions
\begin{align}
\rho_{m_1,m_2}^{m_1',m_2'}(0) &= \frac{1}{2^N}\left[\binom{2s}{m_1}\binom{2s}{m_2}\binom{2s}{m_1'}\binom{2s}{m_2'}\right]^{\frac{1}{2}}.
\end{align}
As a first result we find that the signal $\braket{S_z}$ is identical to the one in Eq.~\eqref{signal}. On the other hand, the standard deviation of the signal is drastically changed: at the point where the signal vanishes (optimal detection point) the variance is
\begin{align}
\braket{S_z^2} &= \frac{N}{4}e^{-\gamma_d\tau}\cosh(\gamma_d\tau).
\end{align}
The significant reduction in the detection quantum noise is immediately apparent by comparing the result's scaling with $N$ to the one in Eq.~\eqref{variance}. Furthermore, \fref{fig5}(b) clearly shows how the noise cancelation mechanism leads to an extremely reduced quantum noise exactly at the points of maximum signal slope where previously the noise was maximized. The frequency detection sensitivity is subsequently enhanced
\begin{align} \label{sens_asym}
\delta\Omega &= \frac{\sqrt{\cosh(\gamma_d\tau)}}{\tau\sqrt{N}}.
\end{align}
The optimal interrogation time is as before around $\tau_{opt}\approx 2/\gamma_d$. However, the result is remarkable in that the optimal sensitivity recovers the $1/\sqrt{N}$ scaling as in the case of the noise-free Ramsey procedure. After numerical minimization with respect to $\tau$, the optimal sensitivity is
\begin{align}
\min_\tau[\delta\Omega] &\approx \frac{0.969}{\sqrt{N}}\gamma_d.
\end{align}

\noindent Note, that the result obtained above can be exactly reproduced by inverting the sign of the phase instead of the detuning in one of the arms of the twin beam interrogation scheme (see Appendix B). This leaves the free Hamiltonian unchanged but yields a modified dissipator describing collective dephasing with the operator $S_z^{(1)}-S_z^{(2)}$, which is the approach used in Ref.~\cite{dorner2012quantum}.

\section{Noise induced limits in Rabi spectroscopy}

Another method routinely used in quantum metrology is Rabi spectroscopy where the population excited during an attempted $\pi$-pulse is monitored against the atom-laser detuning. To model this procedure we introduce a coherent driving laser described by the Hamiltonian from Eq. \eqref{H_laser}. The reversible dynamics of the system are then subject to both the free Hamiltonian $H_0$ and $H_l$. The condition for a perfect population inversion pulse on resonance is that the pulse area equals $\pi$ which requires a pulse duration of $\tau=\pi/2\eta$. The frequency sensitivity is then extracted from the detected population signal at the points of the steepest signal to the left and to the right of the resonance.

\begin{figure}[t]
\includegraphics[width=.8\columnwidth]{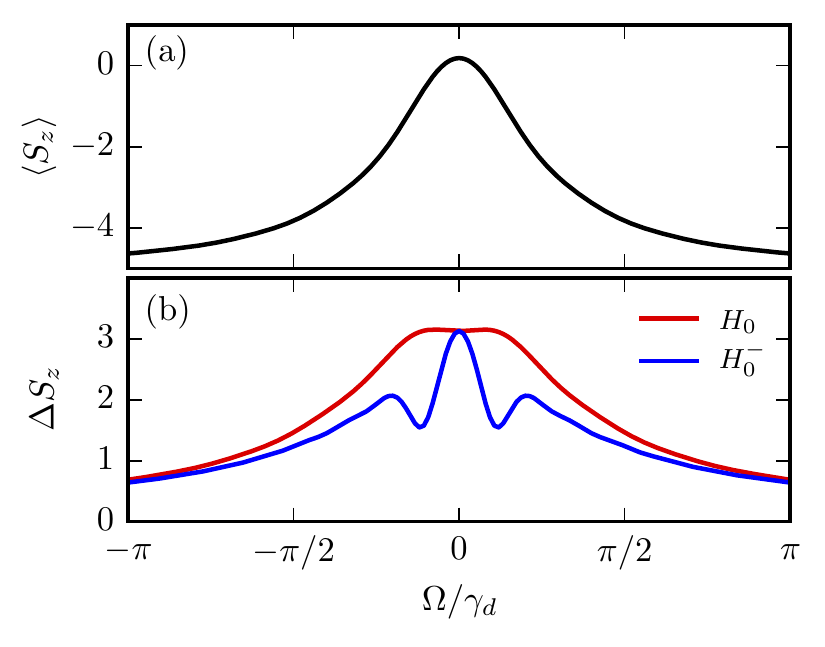}
\caption{\emph{Rabi signal and its quantum standard deviation}. (a) The resonance curve (signal) and (b) the standard deviation which is again exhibiting a significant reduction at the steepest parts to the left and to the right of the resonance when addressing a split ensemble with opposite detunings. The ensemble size is $N=10$ and we chose a sufficiently small driving of $\eta=0.2\gamma_d$ in order to minimize power broadening effects.}
\label{fig6}
\end{figure}

\begin{figure}[ht]
\includegraphics[width=.8\columnwidth]{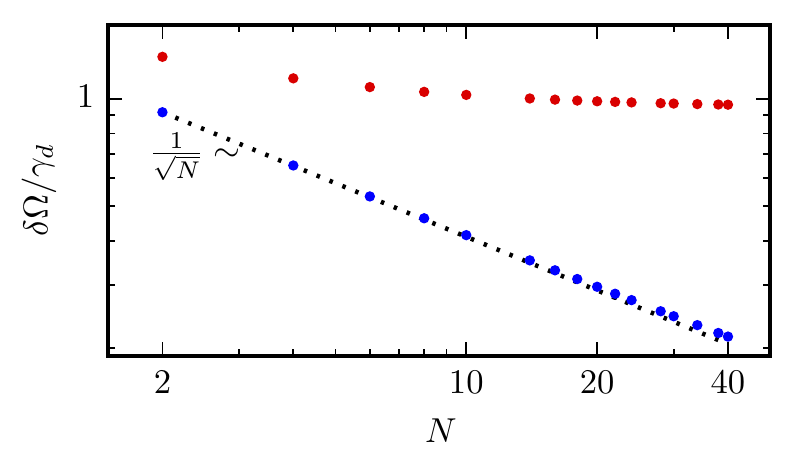}
\caption{\emph{Scaling of the Rabi signal sensitivity with $N$}. While for standard addressing the sensitivity begins to saturate already for $N\sim 20$, when employing opposite detunings the signal sensitivity approximately scales with $1/\sqrt{N}$. An exact fit returns a scaling with $N^{-0.483}$. The driving strength is fixed to $\eta=\gamma_d$.}
\label{fig7}
\end{figure}

We proceed with numerical investigations focused on describing and overcoming the effect of phase noise during the Rabi procedure. The system Hamiltonian for standard Rabi spectroscopy reads
\begin{align}
H_0+H_l &= \Omega S_z + 2\eta S_x,
\end{align}
while for the suppression of noise we will employ the asymmetric free Hamiltonian $H_0^-$ defined in Eq. \eqref{H_asym}. Similarly to the case of Ramsey spectroscopy, we have illustrated an example of the signal and its standard deviation in~\fref{fig6}; the effect of noise reduction occurs again at the optimal operation point where the change of the signal with the detuning is maximal. We quantify the enhancement by using the definition of the signal sensitivity from Eq.~\eqref{sens} as a figure of merit for the frequency detection sensitivity. The red dots in~\fref{fig7} convey the message that phase noise again leads to a saturation effect when the atom number is increased. The blue dots instead approximately show the recovery of the $1/\sqrt{N}$ scaling.

\section{Conclusions}

We have investigated the effects of both laser phase and amplitude noise on the sensitive detection of frequencies via the Ramsey and Rabi technique. For Ramsey spectroscopy we have found analytical expressions showing a saturation of the minimum sensitivity with the increasing atom number under the action of phase noise. To counteract this effect, we have used an approach which involves the separate interrogation of two sub-ensembles leading to the recovery of the typical standard quantum limited scaling with $1/\sqrt{N}$. This setup is closely related to the one illustrated in Ref.~\cite{dorner2012quantum}, but is also reminiscent of techniques used in recent proposals for counteracting the effect of noise~\cite{caves2010coherent,caves2012evading,heurs2014coherent,hammerer2015trajectories,polzik2010quantum,clerk2013two,vitali2016force} in quantum force detection setups: we describe an 'anti-noise' path obtained via the inclusion of a negatively detuned sub-ensemble playing the role of a negative mass oscillator. For amplitude noise, we have also numerically uncovered a particular scaling of the Ramsey sensitivity with $1/\sqrt{\tau}$ at times larger than the inverse of the amplitude noise characteristic rate $\gamma_a^{-1}$. Numerical investigations suggest that similar results hold for Rabi spectroscopy.

\section{Acknowledgements}
We acknowledge very informative and helpful discussions with Laurin Ostermann on theoretical aspects of this work. Also, we thank Aur\'{e}lien Dantan for input on experimental details of laser noise sources. We acknowledge financial support by the Austrian Science Fund (FWF) within the DK-ALM: W1259-N27 (D.~P.) and a stand-alone project with number P24968-N27 (C.~G.), by DARPA through the QUASAR project (H.~R.), and through JILA under grants JILA-NSF-PFC-1125844 and NSF-PIF-1211914 (J.~S.). Furthermore, we acknowledge the use of the open-source software QuTiP~\cite{johansson2013qutip}.


\section{Appendix}

\subsection{Derivation of the master equation}

\subsubsection{Von Neumann equation with a noisy laser}
The Hamiltonian of a perfectly coherent laser source in a frame at rest is
\begin{align}
H_l &= \eta\sum_j \left(\sigma_j^+ e^{-i\omega_lt} + \sigma_j^- e^{i\omega_lt}\right),
\end{align}
where $\sigma_j^\pm$ is the raising and lowering operator of the $j$-th atom, respectively, and $\sigma_j^x = \sigma_j^+ + \sigma_j^-$. We may now include a noisy phase $\phi(t)$ modelled via the standard phase diffusion theory~\cite{graham1970laserlight} and amplitude noise $\epsilon(t)$ by replacing
\begin{align}
\label{ph_noise_sub}
\omega_lt &~\to~ \omega_lt + \phi(t),
\\
\eta &~\to ~ \eta + \epsilon(t),
\end{align}
respectively. Going into the frame rotating with $\omega_l$, we see that the noisy phase $\phi(t)$ is equivalent to a noisy frequency $\dot{\phi}(t)$, since then the substitution in Eq. \eqref{ph_noise_sub} becomes
\begin{align}
\Omega &~\to~ \Omega + \dot{\phi}(t),
\end{align}
while the noise term in the amplitude remains unchanged. We assume both $\dot{\phi}(t)$ and $\epsilon(t)$ to be white noise processes, i.e.
\begin{align}
\braket{\dot{\phi}(t)\dot{\phi}(t')} &= \gamma_d\delta(t-t'),
\\
\braket{\epsilon(t)\epsilon(t')} &= \gamma_a\delta(t-t').
\end{align}
Hence, when writing the respective von Neumann equations for the density matrix $\rho$,
\begin{align}
\dot{\rho} &= -i[H, \rho] + i \dot{\phi}(t)[S_z, \rho],
\\
\dot{\rho} &= -i[H, \rho] + 2 \epsilon(t)[S_x, \rho],
\end{align}
we can interpret each of them as a Stratonovich multiplicative stochastic differential equation~\cite{gardiner1985handbook,dorner2012quantum}.

\subsubsection{Transformation between Stratonovich and It\^o stochastic differential equations}
Let us now more generally consider a differential equation for a stochastic process in one variable,
\begin{align}
\frac{d}{dt}x(t) &= a(x(t),t)+b(x(t),t)\xi(t),
\end{align}
where $\xi(t)$ is a white noise process.
Interpreting this equation as a Stratonovich (S) differential equation, it can be transformed into It\^o (I) form, where it follows different rules of calculus. The transformation relation between the two formalisms is \cite{gardiner1985handbook}
\begin{align}\label{S_I_trafo}
\text{(S)}~dx(t) &= a(x(t),t)dt+b(x(t),t)dW(t)
\\
\text{(I)}~dx(t) &= \left[a(x(t),t)+\frac{1}{2}b(x(t),t)\partial_x b(x(t),t)\right]dt
\notag \\
&+b(x(t),t)dW(t),
\end{align}
where $dW(t)=\xi(t)dt$ is the Wiener increment of the stochastic variable $\xi(t)$.

For the specific form of a multiplicative stochastic differential equation in the Stratonovich formalism,
\begin{align}
&\text{(S)}~&&dx(t) = a_0 x(t) dt + b_0x(t)dW(t),
\end{align}
where $a_0$ and $b_0$ are constants, the transformation into the It\^o form yields
\begin{align}
&\text{(I)}~dx(t) = \left[a_0x(t)+\frac{1}{2}b_0^2x(t)\right]dt + b_0x(t)dW(t).
\end{align}
If $x(t)$ is a Markov process it is non-anticipating, so it holds that the average $\braket{xdW}=0$. Hence, when performing the average over the above equation, we find
\begin{align} \label{multi_noise_trafo}
d\braket{x(t)} &= \left(a_0+\frac{b_0^2}{2}\right)\braket{x(t)}dt.
\end{align}
This relation is all we need to derive our master equation.

\subsubsection{Master equation}
Writing the von Neumann equation for our density matrix subject to an arbitrary multiplicative white noise $\xi(t)$, we have
\begin{align}
\dot{\rho}(t) &= \left(\boldsymbol{L}_0+\alpha\xi(t)\boldsymbol{L}_1\right)\rho(t),
\end{align}
where the linear operator $\boldsymbol{L}_0$ corresponds to the deterministic part of the process, while $\boldsymbol{L}_1$ describes the action of the noise process on the density matrix. We interpret the differential equation as a Stratonovich stochastic differential equation. Using the identity for multiplicative linear white noise in Eq. \eqref{multi_noise_trafo} we have
\begin{align} \label{complete_noise}
d\rho(t)= \left(\boldsymbol{L}_0dt+\frac{1}{2}\alpha^2\boldsymbol{L}_1^2dt\right)\rho(t),
\end{align}
where we used the fact that $\rho(t)$ is Markovian and remains approximately unchanged when averaging over the noise time scale. Therefore Eq. \eqref{complete_noise} yields our respective master equations by identifying the terms from the von Neumann equations
\begin{align}
\boldsymbol{L}_0\rho &:= -i[H,\rho],
\end{align}
and, for phase noise
\begin{align}
\boldsymbol{L}_1\rho &:= i[S_z,\rho],
\notag \\
\alpha &:=\sqrt{\gamma_d}.
\end{align}
For amplitude noise, on the other hand, we have
\begin{align}
\boldsymbol{L}_1\rho &:= i[S_x,\rho],
\notag \\
\alpha &:=2\sqrt{\gamma_a}.
\end{align}

\subsection{Sensitivity of Ramsey spectroscopy subjected to phase noise}
The details of the calculations performed for Ramsey spectroscopy when only considering phase noise will be illustrated here. To this end we will perform a general calculation for a split ensemble addressed with two different detunings $\Omega_1$ and $\Omega_2$, respectively. We then recover the standard case by setting $\Omega\equiv\Omega_1=\Omega_2$ or the enhanced case by setting $\Omega\equiv\Omega_1=-\Omega_2$. As we have shown in the paper, the derivative of each density matrix element is only proportional to the density matrix element, such that we find the general solutions
\begin{align}
\rho_{m_1,m_2}^{m_1',m_2'}(t) &= \rho_{m_1,m_2}^{m_1',m_2'}(0)e^{i\Omega_1t(m_1'-m_1)}e^{i\Omega_2t(m_2'-m_2)}
\notag \\
&\times e^{-\gamma_dt/2\left(m_1'+m_2'-(m_1+m_2)\right)^2}.
\end{align}
In order to compute the signals, we will first consider the following: the prefect $\pi/2$-pulses can be modeled as rotations of the collective Bloch vector about the $y$-axis. The Pauli spin operators are generators of such rotations such that we can write them as
\begin{align}
R_y &= e^{iS_y\pi/2},
\end{align}
where $S_y=\sum_i\sigma_i^y/2$. The final state after the second $\pi/2$-pulse after a free time evolution for a period $\tau$ is then described by the density operator $\rho_f=R_y\rho(\tau)R_y^\dag$. The expectation value of the population inversion at this point is
\begin{align}
\braket{S_z}_f &= \text{tr}\left(S_z\rho_f\right) = \text{tr}\left(R_y^\dag S_z R_y\rho(\tau)\right) = \braket{S_x}_\tau,
\end{align}
where we used the invariance of the trace under cyclic permutations. The signal after the second $\pi/2$-pulse is hence equal to the expectation value of the operator $S_x$ after the free time evolution. The same holds for the operator squared, i.e. $\braket{S_z^2}_f=\braket{S_x^2}_\tau$. We can write the $S_x$ operator as
\begin{align}
S_x &= \frac{1}{2}\left(S^++S^-\right) = \frac{1}{2}\left(S_1^+ + S_2^+ + S_1^-+S_2^-\right),
\end{align}
where $S^\pm=\sum_j\sigma_j^\pm$ are the collective atomic raising and lowering operators. The action of the raising and lowering operators for each of the atomic ensembles on a state is
\begin{align}
S_1^\pm& \ket{s,m_1,m_2} =
\notag \\
&= \sqrt{s(s+1)-m_1(m_1\pm 1)}\ket{s,m_1\pm 1,m_2},
\end{align}
and analogously for $S_2^\pm$. Using these matrix elements of the raising and lowering operators, we can compute the expectation values $\braket{S_x}_\tau$ and $\braket{S_x^2}_\tau$. In order to solve the arising sums we also need to use the initial conditions of these expectation values $\braket{S_x}(0) = N/2$, and $\braket{S_x^2}(0) = N^2/4$, respectively. We then acquire the expression for the signal
\begin{align}
\braket{S_z}_f &= \frac{N}{4}e^{-\gamma_d\tau}\left(\cos(\Omega_1\tau)+\cos(\Omega_2\tau)\right) =
\notag \\
&= \frac{N}{2}e^{-\gamma_d\tau}\cos(\Omega\tau),
\end{align}
where in the second line we used the fact that the signal is identical for both cases. The expectation value of the operator squared reads
\begin{widetext}
\begin{align} \label{var}
\braket{S_z^2}_f = \frac{N}{8}\Bigg[\frac{e^{-2\gamma_d\tau}}{2}\left(\left(\frac{N}{2}-1\right)\left(\cos(2\Omega_1\tau)+\cos(2\Omega_2\tau)\right)+N\cos((\Omega_1+\Omega_2)\tau)\right)+\frac{N}{2}\left(\cos((\Omega_1-\Omega_2)\tau)+1\right)+1\Bigg].
\end{align}
\end{widetext}

At this point we will make the distinction between the two cases of identical and opposite detunings. For simplicity, we will already use the fact that the signal sensitivity is minimal where the signal vanishes but its derivative
\begin{align}
\partial_\Omega \braket{S_z}_f &= -\tau\frac{N}{2}e^{-\gamma_d\tau/2}\sin(\Omega\tau)
\end{align}
is extremal, i.e. where
\begin{align}
\Omega\tau &= (2n+1)\frac{\pi}{2},~n\in\mathbb{Z}.
\label{min_cond}
\end{align}
When setting $\Omega_1=\Omega_2=\Omega$ and using the minimization condition from Eq. \eqref{min_cond}, the variance becomes
\begin{align}
\braket{S_z^2}_f &= \frac{N}{8}\left(e^{-2\gamma_d\tau}\left(-N+1\right)+N+1\right) =
\notag \\
&= \frac{N}{4}e^{-\gamma_d\tau}\left(N\sinh(\gamma_d\tau)+\cosh(\gamma_d\tau)\right).
\end{align}
Substituting this and the derivative of the signal into the definition of the signal sensitivity, we find our expression
\begin{align} \label{sens_sym_app}
\delta\Omega &= \frac{\sqrt{N\sinh(\gamma_d\tau)+\cosh(\gamma_d\tau)}}{\tau\sqrt{N}}.
\end{align}

On the other hand, when we set $\Omega_1=-\Omega_2=\Omega$, Eq. \eqref{var} becomes
\begin{align}
\braket{S_z^2}_f = \frac{N}{8}\left(e^{-2\gamma_d\tau}+1\right) = \frac{N}{4}e^{-\gamma_d\tau}\cosh(\gamma_d\tau),
\end{align}
which of course yields the expression for the significantly reduced signal sensitivity,
\begin{align} \label{sens_asym_app}
\delta\Omega &= \frac{\sqrt{\cosh(\gamma_d\tau)}}{\tau\sqrt{N}},
\end{align}
which scales down more favorably with the number of atoms.

\subsubsection{Equivalence to phase conjugation}
The process used to achieve the gain in measurement precision in Ref. \cite{dorner2012quantum} was phase conjugation on one of the sub-ensembles. Conjugating the phase changes the sign of the noise in one of the sub-ensembles, such that we find a Lindblad operator
\begin{align}
\mathcal{L}^-[\rho] &= \frac{\gamma_d}{2}\left(S_z^-\rho S_z^- - (S_z^-)^2\rho - \rho (S_z^-)^2\right),
\end{align}
where $S_z^- = S_z^{(1)} - S_z^{(2)}$. The Hamiltonian, on the other hand, remains unchanged. Following the above calculation with this model yields the slightly different general solutions for the density matrix elements,
\begin{align}
\rho_{m_1,m_2}^{m_1', m_2'} &= \rho_{m_1,m_2}^{m_1',m_2'}(0)e^{i\Omega t(m_1' + m_1 + m_2' + m_2)}
\notag \\
&\times e^{-\gamma_dt/2\left(m_1' - m_2' - (m_1 - m_2)\right)^2}.
\end{align}
With these solutions one can show that we find the same signal as before and the increased sensitivity from Eq. \eqref{sens_asym_app}.
\vfill

%

\end{document}